# A Graph Neural Network with Auxiliary Task Learning for Missing PMU Data Reconstruction

Bo Li, *Member, IEEE,* Zijun Chen, Haiwang Zhong, *Senior Member, IEEE*, Di Cao, *Senior Member, IEEE*, Guangchun Ruan, *Member, IEEE*

*Abstract*—In wide-area measurement systems (WAMS), phasor measurement unit (PMU) measurement is prone to data missingness due to hardware failures, communication delays, and cyber-attacks. Existing data-driven methods are limited by inadaptability to concept drift in power systems, poor robustness under high missing rates, and reliance on the unrealistic assumption of full system observability. Thus, this paper proposes an auxiliary task learning (ATL) method for reconstructing missing PMU data. First, a K-hop graph neural network (GNN) is proposed to enable direct learning on the subgraph consisting of PMU nodes, overcoming the limitation of the incompletely observable system. Then, an auxiliary learning framework consisting of two complementary graph networks is designed for accurate reconstruction: a spatial-temporal GNN extracts spatial-temporal dependencies from PMU data to reconstruct missing values, and another auxiliary GNN utilizes the low-rank property of PMU data to achieve unsupervised online learning. In this way, the low-rank properties of the PMU data are dynamically leveraged across the architecture to ensure robustness and self-adaptation. Numerical results demonstrate the superior offline and online performance of the proposed method under high missing rates and incomplete observability.

NOMENCLATUR

| | |
|---|---|
| WAMS | wide-area measurement system |
| PDC | phasor data concentrator |
| PMU | phasor measurement units |
| DoS | denial-of-service |
| SVD | singular value decomposition |
| GNN | Graph neural network |
| GAN | generative adversarial networks |
| GAT | graph attention network |
| CNN | convolutional neural network |
| ATL | auxiliary task learning |
| SPD | shortest path distance |
| GAT | graph attention network |
| GRU | gated recurrent unit |
| LSTM | long short-term memory |

This work was supported by the Guangxi Natural Science Foundation under Grant 2025GXNSFBA069263. (Corresponding author: Haiwang Zhong.)

Bo Li is with the School of Electrical Engineering, Guangxi University, Nanning 530004, and the Department of Electrical Engineering, Tsinghua University, Beijing 100084, China (e-mail: boli@gxu.edu.cn).
Zijun Chen is with the School of Electrical Engineering, Guangxi University, Nanning 530004, China (e-mail: 2412391008@st.gxu.edu.cn).
Haiwang Zhong is with the Department of Electrical Engineering, Tsinghua University, Beijing 100084, China (e-mail: zhonghw@tsinghua.edu.cn).
Di Cao is with the School of Mechanical and Electrical Engineering, University of Electronic Science and Technology of China, Chengdu 611731, China (e-mail: caodi@uestc.edu.cn).
Guangchun Ruan is with the Laboratory of Information and Decision Systems, Massachusetts Institute of Technology, Cambridge, MA 02139 USA (e-mail: gruan@ieee.org )

## I. INTRODUCTION

Wide-area measurement systems (WAMS) using PMUs and phasor data concentrators (PDCs) in a hierarchical structure for monitoring, control, and protection of modern power systems [1], [2]. State estimation [3], wide-area fault location [4], and voltage stability assessment [5], play a critical role in the energy management system, which highly relies on accurate and real-time state variables, i.e., bus voltage magnitudes and phase angles, from WAMS. With thousands of PMUs deployed in power systems, the exchange of data involves an enormous amount of real-time data flow. In practice, PMU data often inevitably contain missing observations due to measurement uncertainties, including communication errors, cyberattacks, and other factors. For instance, approximately 10% to 30% of anomalous or missing PMU measurements occur during daily operations in China Southern Power Grid [2], [6]. These corrupted data significantly degrade the security and stability of the system. Therefore, missing PMU data reconstruction approaches against corrupted fragments in WAMS play a significant role in modern power systems.

Some studies have proposed methods tolerant of missing values in specific tasks [7], [8], [9], [10], [11]. These studies focus on particular applications and thus have limited applicability. Consequently, research is increasingly focused on a data-centric perspective for missing data reconstruction to accommodate diverse domains. The PMU missing data reconstruction methods can be classified into two classes: singular value decomposition (SVD) based methods and data-driven methods. Since PMU measurements are sampled at synchronized time instants, and the measurements of nearby PMUs are correlated through the power system topology, the high-dimensional PMU data exhibit a coherence property [12], [13]. Currently, numerous studies have been dedicated to SVD-based data reconstruction methods, which aim to uncover such a coherence property by leveraging the rank of the PMU measurements matrix [14], [15], [16]. However, PMU measurements generally comprise the superposition of an ideal low-rank matrix and various types of noise. Consequently, the accuracy of the SVD-based methods' low-rank matrix results may not be satisfactory.

The other class is emerging data-driven solutions, which typically reconstruct data by utilizing the statistical characteristics and spatial-temporal correlations of PMU data. The long short-term memory (LSTM) and gated recurrent unit (GRU) models are employed to achieve relatively accurate restoration by learning patterns from historical measurement series [17], [18]. However, since sequence-based models require complete sequence data, the absence of historical measurements due to PMU failures or denial-of-service (DoS) attacks may degrade their prediction performance. Compared to sequence-based models, spatial-based models exhibit more stable performance in missing value reconstruction tasks. A missing data reconstruction method based on a convolutional neural network (CNN) is proposed to perform spatial statistical imputation and refine results [19]. A GNN-based model is proposed to recover missing features by extracting topological patterns from PMU measurements [20]. Generative adversarial networks (GANs) provide a generative paradigm that utilizes a minimax game to minimize the distributional error between generated and real data. In the context of significant data corruption, reference [21] designs a robust GAN to enhance the credibility of generated data by neglecting invalid consecutive data. By integrating graph-convolutional layers, a novel graph-convolutional adversarial network is introduced in [22] to capture the topological information of the grid. To enhance robustness, a hybridized random learning GAN model is employed to train and integrate multiple generators for producing a stable aggregated output [23].

However, these data-driven methods have some limitations. First, existing methods inadequately incorporate topology and electrical parameters [4], [24]. Although some studies attempt to integrate topology by employing the GNN, they are forced to compromise accuracy due to the limited PMU coverage. Specifically, the limited PMU coverage results in a mismatch between the topology and the incomplete PMU data, which hinders the GNN from processing PMU data according to the topology. While the state estimation techniques could theoretically estimate the state of those nodes without PMU [25], [26], they become infeasible under high missing data rates. To address this mismatch, some studies treat unmonitored nodes as missing values in the GNN [2], [22], or pre-impute their values using a Gaussian mixture model [27]. These strategies establish a strict one-to-one correspondence between topology and PMU data by setting initial values for unmonitored nodes, while they inevitably introduce errors into the GNN input. Second, existing data-driven methods often ignore exploring the intrinsic approximately low-rank property of PMU data. This characteristic can serve as prior knowledge that reflects the low-dimensional structure of the PMU data. However, such an approximately low-rank property poses a challenge for knowledge integration techniques, such as multi-task learning or transfer learning [1], [28]. Since PMU data does not form a strictly low-rank matrix, the objective of rank minimization may lead to a gradient conflict with the data reconstruction objective, which can further result in negative transfer. Third, the offline-trained models cannot adapt to the concept drift caused by routine grid variations, including topology reconfigurations and load fluctuations [29]. This concept drift undermines the robustness of these models in real-world grids. Overall, there is a lack of a coordinated framework that can integrate low-rank global constraints with data-driven learning of spatial-temporal patterns.

To address these research gaps, this paper proposes an auxiliary task learning (ATL) framework for reconstructing missing PMU data. A K-hop GNN is proposed to enable direct message passing among PMU-equipped nodes. It overcomes the limitation of incompletely observable systems. Inspired by low-rank matrix imputation methods [12],[14], an unsupervised low-rank matrix imputation model based on GNN is pre-trained as an auxiliary task model. The missing data reconstruction task is implemented through a spatial-temporal GNN named ST-Net. The ST-Net integrates the prior knowledge provided by the auxiliary model through a Reinforced GRU component named GRU-R for robust reconstruction. The main contributions are as follows:

(1) A K-hop GCN architecture named PMU-Enabled K-hop GNN (PEK-GCN) is designed to address the limited coverage of PMU. The PEK-GCN constructs a high-order adjacency matrix for direct info-exchange among PMU-equipped nodes. It can operate directly in incompletely observable systems.

(2) An ATL framework is introduced for prior knowledge integration, which comprises an unsupervised low-rank imputation model as the auxiliary task and a model for reconstructing missing data. The auxiliary task extracts the low-rank property of PMU data to guide the reconstruction process and enables online self-adaptation.

(3) A spatial-temporal GNN architecture named ST-Net is proposed to achieve the missing PMU data reconstruction. The information extracted by the ST-Net is integrated with the low-rank matrix from the auxiliary model. This fusion of information ensures that the reconstruction fully exploits the low-rank property inherent in PMU data.

The rest of the paper is organized as follows. Section II introduces the definition of the PMU data reconstruction task and the general framework of the proposed method. The details of PEK-GCN are introduced in Section III. In Section IV, the proposed data-driven low-rank matrix imputation model and the improved loss function are elaborated. The missing value reconstruction framework is introduced in Section V. In Section VI, the effectiveness of the proposed method is validated by a series of simulations. Section VII concludes this article.

## II. PROBLEM FORMULATION

The power network is modeled by an undirected graph $G(V,E)$, where $V$ and $E$ represent the sets of buses and branches, respectively. Matrix $\mathbf{A}$ is used to define the adjacency matrix of the undirected graph. If bus $i$ and bus $j$ are directly connected by a transmission line, then $a_{ij} = a_{ji} = 1$. Due to the investment cost limitations of PMU, the locations of PMU are not always installed on each bus. Suppose $V^M \in V$ as the locations of PMU. PMU measures the complex voltage phasor of its location bus at each discrete sampling instance. Denote $\mathbf{X} = \{x_{m,t}\} \in \mathbb{R}^{m \times t}$ as the PMU measurement matrix without missing data and communication latency. $m = n \times s$ denotes the measurements on the $n$-th channel during the $s$-th dimension PMU data. Denote $\hat{\mathbf{X}}$ and $\tilde{\mathbf{X}}$ as the observation matrix and the reconstructed matrix, respectively. The mask matrix $\mathbf{M} \in \{0,1\}^{m \times t}$ is used to identify the observability of $\mathbf{X}$. If $x_i$ is observed, then $m_i = 1$.

For certain reasons, such as a DoS attack, extreme weather, or other PMU outages, the dispatching center only receives a subset of PMU measurements. There exist complex missing patterns in PMU measurement data, as shown in Fig. 1, including: (1) random missing values; (2) missing data from a single PMU-equipped node over consecutive time intervals; (3) missing data from multiple PMU-equipped nodes within a single time interval; (4) missing data from various PMU-equipped nodes over consecutive time intervals (block-wise missing). Therefore, the missing data reconstruction can be regarded as a typical multi-node spatial-temporal matrix prediction problem, i.e., predicting the most likely PMU measurement (e.g., voltage and phase angle) in the multiple time steps given the previous $K$ PMU observations. The reconstructed matrix $\mathbf{X}'$ comprises generated data $\tilde{\mathbf{X}}$ provided by the proposed missing data reconstruction method and originally observed data. Its composition is formulated as follows:

$$\mathbf{X}' = \mathbf{M} \odot \hat{\mathbf{X}} + (1-\mathbf{M}) \odot \tilde{\mathbf{X}} \quad (1)$$

where $\odot$ represents the element-wise multiplication.

Current methods focus on reconstructing missing data at the current time, given the previous $K$ time steps of PMU observations. However, if PMU experiences over a long period, current methods do not guarantee. In this paper, an auxiliary task learning model is proposed to reconstruct the missing data. The proposed method can effectively handle scenarios with missing data over extended periods. Moreover, this method could reconstruct missing data with consecutive missing data.

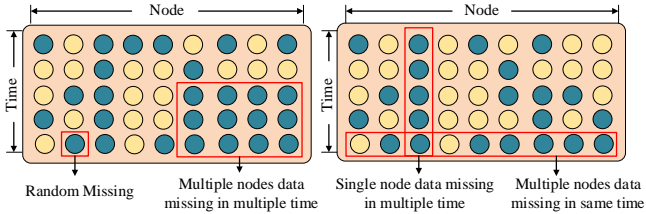

Fig. 1. Missing patterns in PMU measurement data

The overarching concept of this paper is shown in Fig. 2. The raw PMU measurement data are first organized into a time-series observation matrix $\hat{\mathbf{X}}$. $\hat{\mathbf{X}}$ is then mapped onto the physical topology of the power system to construct a graph-structured dataset $\vartheta(\hat{\mathbf{X}}, \mathbf{A}, \mathbf{Y})$. The $\mathbf{Y}$ represents the impedance matrix. Subsequently, the graph dataset $\vartheta$ is fed into two interconnected models: the first model is the self-supervised low-rank matrix imputation model, which serves as the auxiliary task. It enforces the learning of a low-rank prior and outputs the result matrix $\mathbf{L}$ that satisfies the low-rank property. After the calculation of the low-rank imputation result, $\vartheta$ is also processed by the missing data reconstruction model. The reconstruction model extracts spatial-temporal features from $\hat{\mathbf{X}}$ and learns the physical parameters embedded in the impedance matrix $\mathbf{Y}$. The low-rank matrix $\mathbf{L}$ acts as prior knowledge and is input into a middle layer of the missing data reconstruction model to guide the reconstruction process. The features extracted by the missing data reconstruction model are refined by the low-rank prior $\mathbf{L}$ via a learnable weighting mechanism and produce the generated matrix $\tilde{\mathbf{X}}$. The reconstructed matrix $\mathbf{X}'$ is obtained by replacing the missing entries in the original matrix with the corresponding values from the reconstructed matrix $\tilde{\mathbf{X}}$, while keeping the observed entries unchanged, as guided by the mask matrix $\mathbf{M}$.

## III. K-HOP GRAPH NEURAL NETWORK

The limited coverage of PMUs in power systems creates a mismatch between the available node data and the underlying grid topology when applying GNNs. Forcing data completeness through pre-imputation introduces errors into the GNN input, while using a subgraph constructed from available nodes to represent the system topology results in a sparse adjacent matrix that violates the assumption of local connectivity. The emergence of K-hop GNN frameworks provides a new paradigm to overcome this limitation. K-hop GNN aggregates information from multi-hop neighbors (nodes beyond immediate adjacency), enabling the capture of long-range graph patterns. Existing K-hop methods, such as KP-GNN, demonstrate superior performance in complex tasks [30]. Inspired by this approach, a K-hop GNN named PEK-GCN is designed for incompletely observable power systems. By aggregating information from distant nodes within K hops, the PEK-GCN can establish virtual higher-order connections between PMU-equipped nodes, which enables effective information exchange and the capture of long-range dependencies.

The PEK-GCN is a K-hop GNN that captures features from high-order neighbors through diffusion operators. By employing multiple shortest path distance (SPD) kernels as diffusion operators, the PEK-GCN can expand the scope of information aggregation to higher-order neighboring nodes and avoid interference from nodes without PMUs. The $k-$hop neighbors are defined as those nodes with the shortest distance $k$ from the node $i$. According to this definition, the PEK-GCN can be defined as follows:

$$\Theta_k = \tilde{\mathbf{A}}_k \mathbf{X} \quad (2)$$

$$\mathbf{Z} = \sigma\left(\left[\tilde{\mathbf{D}}_1^{-1/2}\tilde{\mathbf{W}}_1\tilde{\mathbf{D}}_1^{-1/2}\Theta_1,...,\tilde{\mathbf{D}}_k^{-1/2}\tilde{\mathbf{W}}_k\tilde{\mathbf{D}}_k^{-1/2}\Theta_k\right]\right) \quad (3)$$

$$\mathbf{H} = \mathbf{W}\mathbf{Z} \quad (4)$$

The high-order adjacency matrix with self-loops is defined as $\tilde{\mathbf{A}}_k$. It is constructed by identifying PMU-equipped nodes among the $k-$hop neighbors of each node and establishing higher-order connection relationships accordingly. According to the $k-$hop adjacency matrix, the model pre-computes a set of diffusion operators $\Theta_1, \Theta_2, ..., \Theta_k$ in Equation (2). They enable the GNN to propagate information across different hop distances in the graph, focusing on the connections involving PMU-equipped nodes. In Equation (3), the PEK-GCN employs these pre-computed diffusion operators to perform a hierarchical graph convolution process. Where $\mathbf{D}_k$ is the degree matrix with self-loops corresponding to $\tilde{\mathbf{A}}_k$; $\mathbf{W}, \mathbf{W}_1, \mathbf{W}_2...\mathbf{W}_k$ are a series of learnable parameter matrices; $\sigma$ represents an activation function. Each diffusion operator is applied to the graph data, and the resulting graph convolution outputs from different diffusion operators are concatenated.

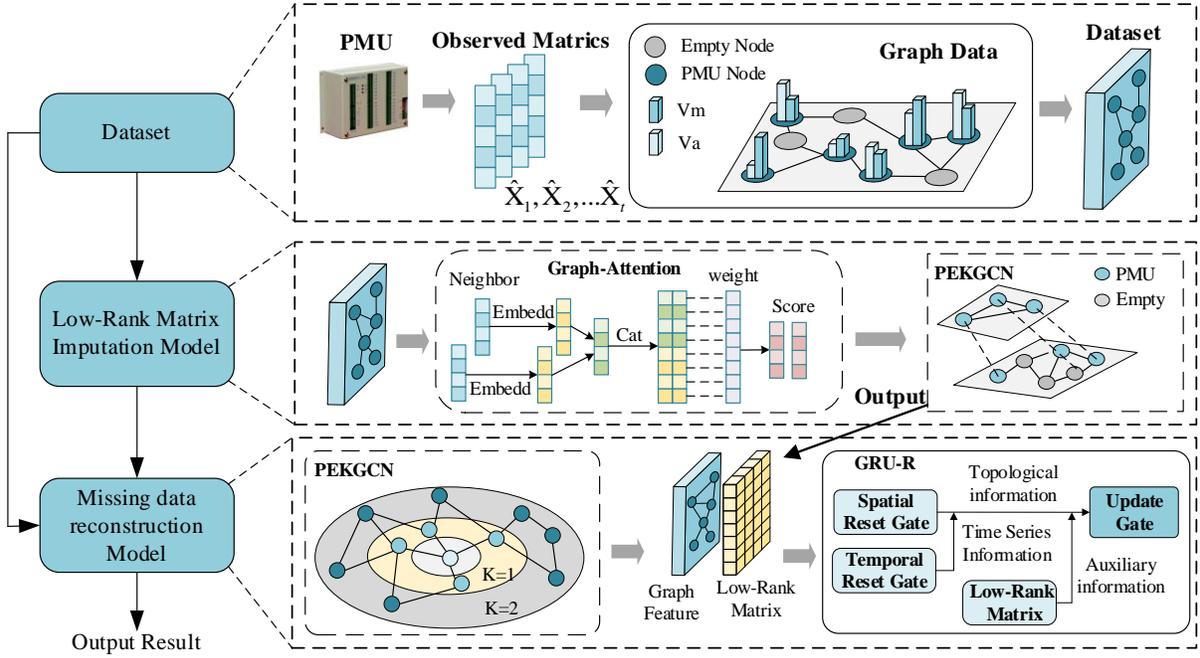

Fig. 2. The complete flowchart of the proposed auxiliary task learning framework

together to form a tensor $\mathbf{Z}$. The tensor $\mathbf{Z}$ includes the information extracted from different hop distances. In Equation (4), the model applies a linear layer to the tensor $\mathbf{Z}$ to fuse the spatial information extracted from different layers and output the result $\mathbf{H}$. By hierarchically aggregating information corresponding to nodes at different hop distances, PEK-GCN avoids the influence of "empty" nodes. Moreover, different diffusion operators correspond to different perception ranges, and the additional information enhances the expressive capacity of the model.

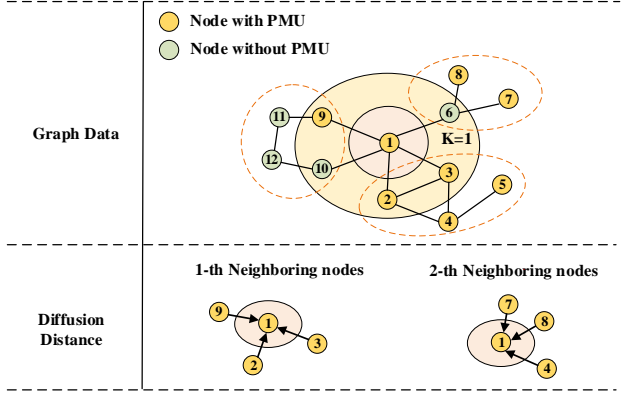

Fig. 3. Schematic diagram of the principle of PEK-GCN.

Fig. 3 illustrates the operating principle of the PEK-GCN. The maximum processing hop count of PEK-GCN is set to 2. Taking node 1 as an example, the first diffusion operator aggregates information from the node's immediate neighbors. It collects data from nodes in the 1-hop range, such as nodes 2, 3, and 9. The 2nd diffusion operator extends the aggregation to a wider range. For node 1, in the second diffusion, it can incorporate information from nodes such as 4, 7, and 8, which belong to its 2-hop neighborhood. For PMU-equipped nodes that exceed the maximum diffusion distance (such as node 5), node 1 won't exchange information with it. By establishing high-order connections among PMU-equipped nodes, the PEK-GCN provides a practical solution to limited PMU coverage that can operate directly in the system with incomplete observability.

IV. LOW-RANK MATRIX IMPUTATION MODEL

In this section, a GNN-based model is constructed to achieve the low-rank matrix imputation task. To adapt to the data-driven structure, an improved nuclear norm is proposed as the loss function to facilitate the optimization process.

A. Model Architecture

For the low-rank matrix reconstruction problem, the most prevalent approach is to minimize the rank of a matrix with missing values to obtain an optimal low-rank matrix. Multiple studies have demonstrated that optimization models can achieve satisfactory results under a low missing rate. The low-rank matrix reconstruction method based on an optimization model can be formulated as [14], [31]:

$$\min \ \mathrm{Rank}(\mathbf{X}) \qquad (5)$$
$$\mathrm{st.}\ P_\Omega(\mathbf{X}) = P_\Omega(\mathbf{M}) \qquad (6)$$

Equation (5) defines the objective function as minimizing the rank of the input matrix, aiming to reconstruct the input matrix based on the low-rank assumption. Equation (6) constrains the non-missing parts to remain unchanged before and after optimization, where the $P_\Omega$ is an orthogonal projection onto the subspace of matrices supported on $\Omega$. However, when a high proportion of data is missing, the performance of the optimization model may be unsatisfactory. Nevertheless, these methods reveal the feasibility of extracting rank-based prior knowledge from matrices.

From a structural perspective, the neural network makes it hard to guarantee the low-rankness of the output matrix. The main reason is that neural networks lack the mechanism to encode the structural prior of intrinsic similarity among row or column vectors. A low-rank matrix essentially implies high linear correlations between its rows and columns. But the transformations in neural networks (such as CNN) are local and dense. These transformations obscure the structural characteristics of the vectors.

The GNN offers a novel paradigm to address this challenge. In low-rank matrix reconstruction, the message-passing mechanism of graph convolution performs Laplacian smoothing over the graph. This operation enhances the linear dependence among features from connected nodes, guiding the entire graph data toward a low-rank structure. As shown in Equation (7), the node feature matrix is multiplied by the adjacency matrix $\tilde{\mathbf{A}}$ and the degree matrix $\tilde{\mathbf{D}}$ to aggregate information between each node and its neighbors. Then, the learnable matrix $\mathbf{W}^l, \mathbf{B}^l$ is applied to linearly transform the aggregated features, producing final node representations. Since vertices in the same cluster tend to be densely connected, the smoothing operation makes their features similar.

$$\mathbf{X}^{l+1} = \sigma(\tilde{\mathbf{D}}^{-\frac{1}{2}}\tilde{\mathbf{A}}\tilde{\mathbf{D}}^{-\frac{1}{2}}\mathbf{X}^l\mathbf{W}^l + \mathbf{B}^l) \qquad (7)$$

Given the Laplacian smoothing property of GNN, this paper proposes a low-rank matrix reconstruction model. The model employs PEK-GCN and a graph attention network (GAT) to perform smoothing operations across nodes. It is worth noting that the GAT employed in this paper also adopts the K-hop GCN strategy. When computing graph attention and performing attention weighting, GAT only operates on nodes equipped with PMUs. Additionally, the proposed model utilizes a self-attention mechanism to capture global dependencies between row vectors. The low-rank matrix reconstruction process mainly consists of Equations (8-11):

$$\mathbf{A}_\alpha = \text{soft max}\left(\frac{\mathbf{X} \cdot \mathbf{W}_q \cdot (\mathbf{X} \cdot \mathbf{W}_k)^T}{\sqrt{d_k}}\right) \cdot (\mathbf{X} \cdot \mathbf{W}_v) \qquad (8)$$

$$\alpha_{i,j} = \frac{\exp\left(\sigma\left(a^T\left(\left[\mathbf{W}_1\mathbf{X}_i \| \mathbf{W}_2\mathbf{X}_j\right]\right)\right)\right)}{\sum_{j \in N_i^k} \sigma\left(a^T\left(\left[\mathbf{W}_1\mathbf{X}_i \| \mathbf{W}_2\mathbf{X}_j\right]\right)\right)}, i \in \mathbf{V}^M \qquad (9)$$

$$\mathbf{H}_i^{(1)} = \sigma\left(\sum_{j \in N_k} \alpha_{i,j}\mathbf{W}_i\mathbf{X}_i\right) \qquad (10)$$

$$\mathbf{H}_i^{(2)} = \sigma\left(\mathbf{W}_i^{(l)}\mathbf{H}_i^{(2)} + \sum_{j \in N_k} \frac{1}{N}\mathbf{W}_j^{(l)}\mathbf{H}_j^{(1)}\right) \qquad (11)$$

From the perspective of matrix analysis, the proposed model achieves information exchange between row and column vectors, respectively, through Self-Attention mechanisms and GNN. As shown in Equation (8), the proposed model applies a self-attention mechanism to the input matrix. The self-attention mechanism uses three learnable matrices $\mathbf{W}_k, \mathbf{W}_v, \mathbf{W}_q$ to generate keys, values, and queries through multiplication. The key is multiplied by the query, and the result is divided by a scaling factor $\sqrt{d_k}$ ($d_k$ is the dimension of the feature vector) to obtain the attention score. The attention scores are then multiplied by the value matrix to produce the final results. The self-attention establishes coupling relationships between different row vectors. It effectively accomplishes row-wise information exchange in the input matrix.

The PEK-GCN and GAT are employed to enforce smoothness among connected nodes, as shown in Equations (9-11). In Equation (9), the feature $\mathbf{X}_j$ of the node $i$ and the feature $\mathbf{X}_j$ of its $k$-th neighboring node $j$ are combined into a vector (the $N_j^k$ represents the set of all $k$-th neighbors of node $i$), multiplied by learnable parameters $\mathbf{W}_1, \mathbf{W}_1, a$ to get feature similarity, and then normalized for final graph attention scores $\alpha_{i,j}$. In Equation (10), the node feature $\mathbf{X}_i$ is updated to the result $\mathbf{H}_i^{(1)}$ by aggregating information from all $k$-th neighbors based on the graph attention scores $\alpha$. In Equation (11), the proposed model employs PEK-GCN to further facilitate node feature exchange. Equation (11) has the same meaning as Equations (2-4). Therefore, PEK-GCN and GAT perform information passing operations among different node features (i.e., column vectors), which further facilitates the convergence of the matrix to a low-rank solution.

*B. Loss Function Design*

In a low-rank matrix imputation model based on an optimization method, the nuclear norm of the matrix is usually employed as the objective function rather than the rank of the matrix, since directly optimizing the rank of the matrix is an NP-hard problem [14]. The nuclear norm is the sum of all the singular values of a matrix. For a matrix $\mathbf{X} \in \mathbb{R}^{m \times n}$, its singular value decomposition can be described as: $\mathbf{X} = \mathbf{U}\mathbf{S}\mathbf{V}^T$, where $\mathbf{U} \in \mathbb{R}^{m \times n}$ and $\mathbf{V} \in \mathbb{R}^{m \times n}$ are orthogonal matrices, and the $\mathbf{S} \in \mathbb{R}^{m \times n}$ is a diagonal matrix composed of non-negative singular values $\sigma_1 \geq \sigma_2 \geq .... \geq \sigma_r > 0$. Thus, the nuclear norm of the matrix is described as:

$$\min \|\mathbf{A}\|_* = \min \sum_{i=1}^{r} \sigma_i(\mathbf{A}) \qquad (12)$$

When the nuclear norm serves as the objective function in an optimization model, it ensures the low-rank property of the output matrix by controlling the distribution of singular values. However, since neural networks inherently yield approximately optimal solutions rather than globally optimal solutions, the convergence of this objective function within a neural network framework is a challenge. In the optimization model, when the optimal solution $\mathbf{A}^*$ is obtained, all singular values $\sigma_r^*$ have reached their optimal states, with some of the smaller ones approaching zero. Consequently, in the optimization model, the $\|\mathbf{A}^*\|_* < \|\mathbf{A}\|_*$ can ensure $\text{Rank}(\mathbf{A}^*) < \text{Rank}(\mathbf{A})$. However, in the neural network framework, this conclusion may not be valid. This phenomenon arises from biased parameter updates resulting from large disparities in singular value magnitudes, which lead to gradient imbalances. Large singular values dominate gradient optimization, rapidly reducing the objective value but failing to effectively lower the matrix rank. Consequently, small singular values struggle to converge to zero, trapping the model in a local optimum.

To address this issue, a regularized nuclear norm is proposed to replace the nuclear norm:

$$\|\mathbf{A}_{n\times m}\|_{\text{Norm}} = \sum_{i=1}^{\min(n,m)} log(\sigma_i(\mathbf{A})+\varepsilon) \qquad (13)$$

Equation (13) employs the logarithmic function to regularize the singular values. This regularized nuclear norm reduces the magnitude differences among different singular values. The regularized singular values are denoted as $log(\sigma_i+\varepsilon)$. And the $\varepsilon$ represents a small positive number, which is introduced to prevent the logarithmic function from accepting zero when some singular values are optimized to zero. The gradients corresponding to the nuclear norm and the regularized nuclear norm can be described as:

$$\nabla\|\mathbf{A}\|_* = \sum_{i=1}\frac{\partial L}{\partial \sigma_i}\frac{\partial \sigma_i}{\partial \theta} = \sum_{i=1} 1\cdot\frac{\partial \sigma_i}{\partial \theta} \qquad (14)$$

$$\nabla\|\mathbf{A}\|_{\text{Norm}} = \sum_{i=1}\frac{\partial L}{\partial \sigma_i}\frac{\partial \sigma_i}{\partial \theta} = \sum_{i=1} \frac{1}{\sigma_i+\varepsilon}\cdot\frac{\partial \sigma_i}{\partial \theta} \qquad (15)$$

$\nabla\|\mathbf{A}\|_{\text{Norm}}$ denotes the gradients of singular values with different magnitudes. In this way, each singular value can be optimized fairly, ensuring a stable decrease in the matrix rank during the model optimization process.

## V. PMU Measurement Reconstruction Model

### A. Spatial-Temporal GNN Architecture

Currently, numerous studies have leveraged the low-rank property of PMU measurements from the perspective of optimization models [14], [15], [16]. In contrast, few studies have integrated the low-rank property of the PMU measurement matrix as prior knowledge into data-driven models from the perspective of multi-task learning.

Therefore, an ATL framework is proposed to reconstruct missing data. ATL is a learning strategy that leverages auxiliary tasks to enhance the performance of the primary task [32]. Compared to Multi-Task Learning and Transfer Learning, the collaborative relationships among different tasks in ATL are more flexible. It enables the auxiliary tasks to play a supporting role without interfering with the optimization process of the missing data reconstruction task. In this paper, the proposed ATL framework takes missing value reconstruction as the primary task, and the exploration of matrix low-rank properties serves as an auxiliary task to support missing value reconstruction. The entire framework consists of two components: a spatial-temporal GNN named ST-Net, and an auxiliary model, which has been introduced in Section III.

ST-Net is a spatial-temporal GNN based on the ST-GCN architecture [33]. ST-Net retains the iconic "sandwich" structure of ST-GCN and enhances its performance in the PMU data reconstruction task through several improvements. As shown in Fig. 4, ST-Net integrates the low-rank matrix reconstruction model with two spatial-temporal blocks (ST-blocks). Each ST-block comprises three sequential components: the PEK-GCN for extracting topological features, the GAT for dynamic edge weighting, and the GRU-R for constructing temporal dependencies and fusing low-rank matrix results. This hierarchical architecture enables ST-Net to capture both the static topological relationships and dynamic patterns in power systems.

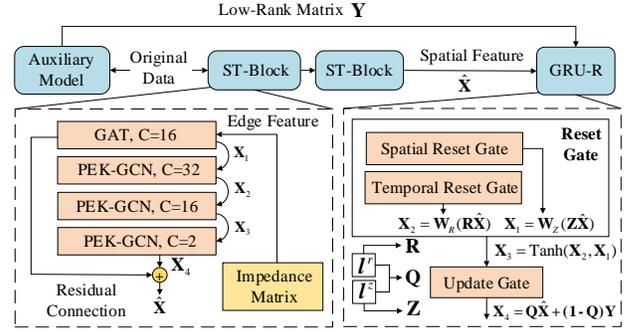

Fig. 4. Schematic diagram of the missing value reconstruction framework

Spatially, ST-Net utilizes PEK-GCN for constructing long-range topological relationships and capturing spatial dependencies among the grid. And ST-Net also uses the GAT as an edge feature enhancer. By encoding impedance matrices into dynamic edge weights, the ST-Net can adaptively learn the key paths. The computational process of GAT can be expressed as follows:

$$e_{m,n} = |(\mathbf{G}_{m,n} + j\cdot\mathbf{B}_{m,n})^{-1}| \qquad (16)$$

$$\alpha_{i,j} = \frac{\exp\left(\sigma\left(a^T\left([\mathbf{W}_1\mathbf{H}_i^{(0)}\|\mathbf{W}_2\mathbf{H}_j^{(0)}\|e_{i,j}]\right)\right)\right)}{\sum_{k\in N_i^k}\sigma\left(a^T\left([\mathbf{W}_1\mathbf{H}_i^{(0)}\|\mathbf{W}_2\mathbf{H}_j^{(0)}\|e_{i,j}]\right)\right)}, \text{ i}\in\mathbf{V}^M \qquad (17)$$

Given that the impedance matrix effectively quantifies the electrical coupling strength between any pair of nodes. The ST-Net obtains the impedance matrix from the admittance matrix $\mathbf{G}_{m,n}+j\cdot\mathbf{B}_{m,n}$. Then, the admittance matrix is inverted, and its modulus values are used as edge features, as in Equation (16). In Equation (17), compared to the attention score calculation in Equation (9), the GAT in ST-Net concatenates edge features $e_{m,n}$ with node features during the computation process. By considering interactions between different nodes, the proposed model can effectively extract spatial and temporal relationships. For instance, the voltage magnitude $V_{i,t}$ of nod $i$ can take into account the influence of the voltage phase angle $\theta_{j,t}$ from other nodes $j$ that have potential associations.

### B. Auxiliary Learning Strategy

The ST-Net extracts information from three key sources: prior knowledge from the auxiliary model, spatial dependencies captured by GNN, and time-series dependencies. To effectively fuse these heterogeneous data streams and properly utilize the prior knowledge provided by the low-rank matrix, the GRU-R architecture is proposed. GRU-R employs a novel gating mechanism to dynamically combine features from different sources.

Specifically, GRU-R splits the reset gate into a temporal reset gate and a spatial reset gate. The temporal reset gate is designed to control the proportion of information provided by PMU measurements at the previous time step; the spatial reset gate is designed to control the proportion of information extracted by PEK-GCN. Update gate serves to maintain the proportion of spatial-temporal dependencies provided by reset gates and the prior feature provided by the auxiliary task.

The calculation processes of GRU-R mainly comprise Equations (18-22):

$$r_t = \exp\{-\max\{0, w_r l_t^r + b_r\}\} \quad (18)$$

$$z_t = \exp\{-\max\{0, w_z l_t^z + b_z\}\} \quad (19)$$

$$q_t = 1 - \exp\{-\max\{0, u_q l_t^r + w_q l_t^z + b_q\}\} \quad (20)$$

$$\hat{h}_t^{(1)} = \tanh\left(w_c \cdot (z_t \cdot \tilde{h}_t^{(0)}), u_c \cdot (r_t \cdot h_{t-1}^{(0)})\right) \quad (21)$$

$$h_t^{(1)} = q_t \cdot \hat{h}_t^r + (1 - q_t) \cdot \hat{h}_t^{(1)} \quad (22)$$

The GRU-R initially calculates the feature missing rates $l_t^r$ and $l_t^z$ at the time $t$. $l_t^r$ represents the proportion of time steps with missing data before time $t$ for node $i$; $l_t^z$ represents the proportion of nodes with missing data among the high-order neighbor nodes with PMUs. As illustrated in Equation (18-19), the model calculates the temporal reset gate $r_t$, spatial reset gate $z_t$, and update gate $q_t$ based on the missing rate $l_t^r$ and $l_t^z$. The $w_r$ and $b_r$ are the learnable parameters belonging to the spatial reset gate. The $w_z$ and $b_z$ are the learnable parameters belonging to the temporal reset gate. The $u_q$, $w_q$, and $b_q$ are learnable parameters belonging to the update gate.

Based on the reset gate and update gate, the spatial feature $\tilde{h}_t^{(0)}$, the temporal feature $h_{t-1}^{(0)}$, and the low-rank matrix result $\hat{h}_t^r$ are retained and updated in Equations (21)-(22). When the $l_t^z$ is larger than the $l_t^r$, the model places greater trust in temporal dependencies rather than spatial dependencies. More information from the previous time step will be retained during the update process, as described in Equation (21). When $l_t^r$ and $l_t^z$ both are relatively large, it signifies a severe lack of spatial-temporal context for the node. In this case, the $r_t$ and $z_t$ in Equation (21) are reduced, while the $q_t$ in Equation (22) is increased. The model then relies more on the information provided by the low-rank matrix, rather than the spatial-temporal information provided by neighboring nodes and historical data. When the $l_t^r$ and $l_t^z$ are both relatively small, the model places greater reliance on the spatial-temporal information.

## VI. SIMULATION RESULT

### A. Data generation

The Databases are generated on the 145-bus case [34]. The sampling frequency is set to 50 Hz. The corresponding nodes in this 145-bus case are selected for data collection based on the classical optimal PMU placement techniques in [35], aiming to achieve N-1 observability. The PMU placement strategy of the test cases is shown in Table I.

TABLE I
NUMBERING OF THE PMU-EQUIPPED NODE

| Case | Node |
|---|---|
| 145-buses Case | 0,1,6,21,26,32,35,36,39,41,45,46,50,57,58,59,60,65,68,71, 72,73,75,116,118,131,138 |

For the 145-bus case, Latin Hypercube Sampling is applied to vary the load levels from 115% down to 85% of the nominal value. A total of 10,000 feasible load conditions are generated, where convergence is verified, and infeasible cases are discarded through power flow calculation. The load conditions are further employed to generate voltage measurements. To ensure the accuracy of the generated data, the voltage data are obtained through transient stability simulations conducted using the Power System Toolbox (PST) [36]. The accuracy of generating synthetic data via time-domain simulation techniques has been validated in multiple existing studies [2], [18], [21].

Transient stability assessment is performed for each load condition after applying three-phase-to-ground short-circuit faults. These faults are cleared within a fault-clearing time of 0.2 seconds. Following fault clearance, bus voltage values are collected over consecutive time intervals. And all bus voltages are maintained within 0.95-1.05 p.u. The collected measurements are structured as matrix data $X \in R^{T \times d}$. $T$ represents the number of time steps, which is set to 8 in this study. $d$ represents which bus the feature is located on. The dataset is split into 70% training sets, 20% testing sets, and 10% validation sets.

To simulate realistic PMU failures and communication losses, the data missing rate in both training and testing sets is uniformly set at 30%, reflecting typical 10%-30% PMU data loss observed in power system operations [2]. Considering non-random factors like multi-period and multi-node data missingness caused by DoS attacks or extreme disasters, consecutive missingness events are imposed on over half of the PMU-equipped nodes. The simulation introduced 30 distinct data missing events on these nodes over 10,000 time steps, with each event lasting for 300 consecutive steps. The overall missing rate in both the training and testing sets reached 60%. Compared to methods that consider only random PMU failure rates [19], [21], [37], the dataset generated in this work exhibits more complex patterns of missing data.

### B. Performance of Proposed Model

Details of the proposed low-rank matrix imputation model and ST-Net are given in Tables II and III. The proposed model is implemented using the PyTorch framework and optimized with the Adam optimizer. The batch size is set to 300. The learning rates are configured as 0.005 for the low-rank matrix imputation model and 0.01 for the ST-Net model.

TABLE II
DETAILS OF THE LOW-RANK MATRIX IMPUTATION MODEL

| Layer | Description |
|---|---|
| 1st GCN Layer. | Hidden Size=2, Activation=ReLu |
| 2nd GCN Layer. | Hidden Size=2, Activation=ReLu |
| Self Attention | Hidden Size=64, Dropout=0.1 |

TABLE III
DETAILS OF THE ST-NET MODEL

| Layer | Description |
|---|---|
| GAT Layer | Hidden Size=16, Activation=ReLu |
| 1st GCN Layer | Hidden Size=32, Activation=ReLu |
| 2nd GCN Layer | Hidden Size=16, Activation=ReLu |
| 3rd GCN Layer | Hidden Size=2, Activation=ReLu Residual from GAT (α=0.3) |
| Batch Normalization | Input Size=2 |
| GRU-R | Hidden Sized=16, Activation=Tanh |

The proposed model is compared with existing methods, including: LSTM [38], GRU [18], GAN [21], self-attention network [39], non-local network [40], and a conventional optimization-based model [14], which formulates the reconstruction task as a constrained rank-minimization problem solved via an optimization solver.

Table IV presents the performance benchmark of the proposed model on the bus-145 test case. The proposed model achieves the lowest Mean Squared Percentage Error (MSPE) and Root Mean Square Error (RMSE) among all compared methods, demonstrating its strong capability in reconstructing missing PMU data in large-scale power systems. The GAN exhibits competitive performance, yet its reconstruction accuracy is slightly lower than that of the proposed method. This gap may stem from the difficulty of GANs in maintaining equilibrium between the generator and discriminator, which can lead to suboptimal convergence. Moreover, GANs' emphasis on distribution matching may overlook the precise reconstruction of individual missing values.

Sequence-based models such as LSTM and SRPF, which rely on temporal structures, exhibit limited performance in reconstructing PMU measurements. This is primarily due to their difficulty in capturing the complex, non-sequential spatial dependencies inherent in large power networks. As a result, they exhibit higher RMSE and MSPE in this task. Similarly, despite attention-based mechanisms like self-attention and non-local networks that can construct long-range dependencies, their performance degrades when irregular missing patterns lead to suboptimal weight allocation. Notably, the optimization model yields the highest error in its result. This result highlights the limitation of convex optimization-based methods in handling highly incomplete data.

The superior performance of the proposed model stems from its auxiliary task learning architecture, where the auxiliary module provides robust, coarse-grained, and low-rank matrix estimates, while the spatial-temporal network extracts fine-grained features. The GRU-R module can effectively fuse these multi-granular information streams and achieve more accurate and stable reconstruction.

TABLE IV
PERFORMANCE BENCHMARKING OF THE PROPOSED MODEL

| Method | MSPE (%) | RMSE |
|---|---|---|
| Proposed Model | **1.056** | **0.01883** |
| GAN | 1.587 | 0.02632 |
| GRU | 2.374 | 0.02954 |
| LSTM | 2.302 | 0.05508 |
| Self-Attention | 2.715 | 0.07139 |
| Non-Local | 2.588 | 0.02838 |
| Optimization Model | 54.13 | 0.67449 |

### C. Performance of Low-Rank Matrix Imputation Model

To evaluate the performance of the auxiliary learning model, SVD is performed on three matrices: the original input matrix, the reconstruction result of the proposed model, and the ground truth matrix. The singular values can precisely reflect the intrinsic structural properties of matrices.

Figure 5 illustrates the distribution of singular values for the three matrices. Since the complex missing patterns exist in the original data, obscure dimensional correlations, all singular values exhibit abnormally high magnitudes. In contrast, both the ground truth matrix and the result of the proposed model exhibit low-rank characteristics: the first singular value accounts for more than 95% of total energy (with subsequent values approaching zero). This phenomenon demonstrates the model's ability to extract low-rank structural components and suppress redundant information. However, the proposed model demonstrates incomplete recovery of the dominant singular value, suggesting a limitation in capturing critical information. Consequently, the auxiliary learning only provides coarse-grained structural guidance rather than fine-grained feature alignment.

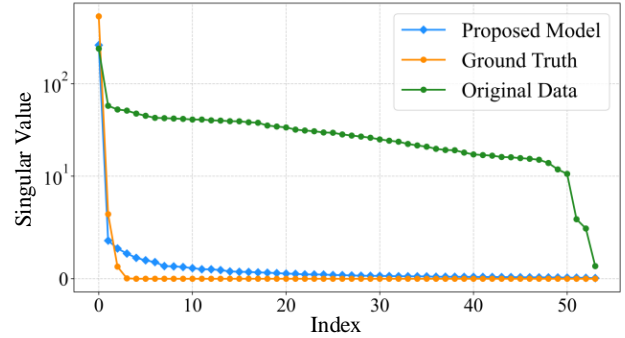

Fig. 5. The optimization effect of the proposed model on singular values

The optimization performance of the proposed improved nuclear norm is compared with that of the standard nuclear norm. As shown in Fig. 6, the proposed model, equipped with an improved kernel norm loss function, achieves superior low-rank matrix recovery. Most of the singular values are effectively suppressed below 1. This suppression effect demonstrates that the improved nuclear norm ensures a more equitable gradient distribution through normalization across singular values, leading to superior rank minimization and stable convergence.

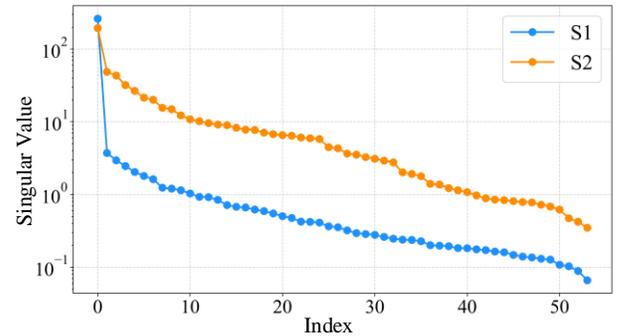

Fig. 6. Comparison of two singular value optimization methods. S1 and S2 represent the optimization performance of the improved nuclear norm and the standard nuclear norm, respectively.

### D. Ablation Experiment

To validate the effectiveness of the proposed approach, a comprehensive ablation study is conducted. The proposed model is compared with its two variants. The variant V1 replaces the PEK-GCN with a standard GNN. It trains on a fully observed system. By disabling the low-rank matrix data pathway, variant V2 prevents the GRU-R from utilizing information provided by auxiliary tasks.

As shown in Fig. 7, the proposed model achieves the best performance across most evaluation metrics, with the sole exception of MSPE. Variant V1 slightly outperforms the MSPE of the proposed model. This discrepancy may be attributed to the fact that variant V1 is trained on fully observed system data. So variant V1 can capture more accurate patterns at certain nodes where more node data is available.

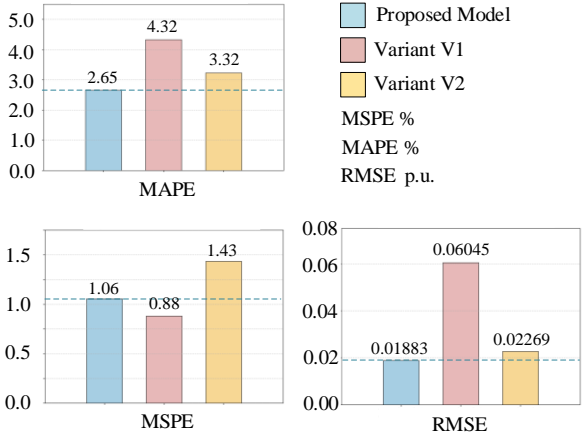

Fig. 7. The performance of the proposed model and its two variants.

The ablation results demonstrate that the accuracy of PEK-GCN under incomplete observability remains comparable to that in fully-observed systems, with no significant degradation in performance. Therefore, the proposed PEK-GCN is confirmed to effectively process incompletely observable PMU data, thereby overcoming the difficulty that GNNs face in operating under limited PMU coverage. Moreover, a significant decline in prediction accuracy is observed in the variant V2. It demonstrates that the proposed auxiliary learning framework makes a significant contribution to overall performance enhancement. Overall, the ablation study confirms that both key modules in the proposed model make contributions to the performance of the prediction task.

*E. Online Performance of The Proposed Model*

This section designs an online prediction framework for our model and compares its performance with the GRU-based online prediction framework in [18] and the GAN-based online prediction framework in [21]. Following the method in Section VI-A, a dataset is collected for online evaluation. The online dataset is collected with a 20-step delay in transient stability assessment compared to the training dataset.

Fig. 8 shows the details of the online prediction framework. Unlike online frameworks that rely on static offline learning, the proposed framework integrates an unsupervised auxiliary task, thus possessing the capability for online learning and adaptation. The model performs real-time inference combined with periodic fine-tuning. After predicting a batch of data, the auxiliary model uses the raw features for unsupervised learning. This process dynamically adjusts the parameters of the auxiliary model. If the updates of the auxiliary model lag behind the performance degradation of the ST-Net, a pseudo-label mechanism is activated. This mechanism trades increased computation time for higher accuracy. Specifically, it uses the outputs of the auxiliary model as supervisory signals to fine-tune the ST-Net and thus achieves online calibration.

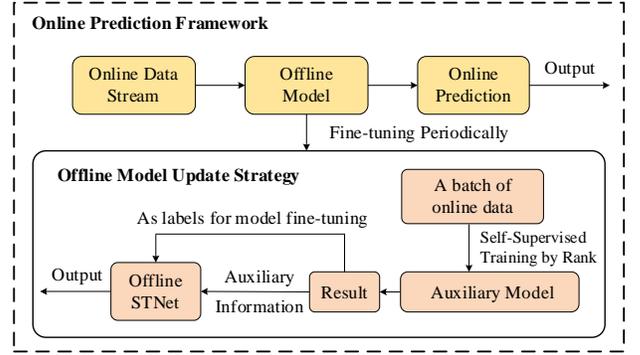

Fig. 8. Flowchart of the proposed online prediction framework.

Table V presents the online performance benchmark of the models. Models with a fine-tuning capability underwent an initial 8,000 steps for online adaptation, and their performance is evaluated over the subsequent 2,000 steps. Models without fine-tuning are evaluated directly over a 2,000-step prediction window. When the voltage distribution pattern changes in the initial stage, both the proposed method and SRPF exhibit a significant decline in performance. In contrast, the performance degradation of the GAN is notably smaller. After a period of unsupervised training, the proposed model's performance gradually recovered and then surpassed that of the GAN. With the subsequent application of pseudo-labels to adjust the parameters of ST-Net, its prediction accuracy further improved, closely approaching the level achieved during offline training. This demonstrates the model's robust online prediction performance.

TABLE V
ONLINE PERFORMANCE BENCHMARKING OF THE PROPOSED MODEL

| Model | MSPE (%) | RMSE |
| --- | --- | --- |
| Initial Stage | 157.228 | 0.28279 |
| Self-Supervised Stage | 6.067 | 0.15331 |
| pseudo-label Stage | **2.952** | **0.03474** |
| GAN | 21.191 | 0.26845 |
| SRPF | 157.367 | 0.30725 |

## VII CONCLUSION

This paper proposes a Spatial-Temporal GNN model based on Auxiliary Task Learning for PMU data reconstruction. The effectiveness of each component is verified through simulations. The model is comparatively evaluated against existing methods in test cases of different scales, showing the best performance under complex missing patterns. Unlike existing methods, it integrates the low-rank property of PMU data via an auxiliary task learning framework to extract more complex feature dependencies. These improvements introduce a new paradigm for reconstructing missing data. The proposed PEK-GCN can operate in incompletely observable systems, which makes it more practical than graph-based methods that require full observability. Numerical results demonstrate the superior offline and online performance of the proposed method under high missing rates and incomplete observability. The effectiveness of components in the proposed model is proven through an ablation study. Future research will aim to develop an efficient K-hop GNN with less computational cost for scalable deployment and to extend the framework to address the communication latency in PMU data streams.

## REFERENCES

[1] Q. Li, C. Ren, R. Zhang, and Y. Xu, "A Multi-Task Learning-Based Approach for Power System Short-Term Voltage Stability Assessment With Missing PMU Data," *IEEE Trans. Autom. Sci. Eng.*, vol. 22, pp. 13187–13197, 2025.

[2] J. J. Q. Yu, D. J. Hill, V. O. K. Li, and Y. Hou, "Synchrophasor Recovery and Prediction: A Graph-Based Deep Learning Approach," *IEEE Internet Things J.*, vol. 6, no. 5, pp. 7348–7359, Oct. 2019.

[3] S. G. Ghiocel et al., "Phasor-Measurement-Based State Estimation for Synchrophasor Data Quality Improvement and Power Transfer Interface Monitoring," *IEEE Trans. Power Syst.*, vol. 29, no. 2, pp. 881–888, Mar. 2014.

[4] W. Li, D. Deka, M. Chertkov, and M. Wang, "Real-Time Faulted Line Localization and PMU Placement in Power Systems Through Convolutional Neural Networks," *IEEE Trans. Power Syst.*, vol. 34, no. 6, pp. 4640–4651, Nov. 2019.

[5] L. Zhu, C. Lu, I. Kamwa, and H. Zeng, "Spatial–Temporal Feature Learning in Smart Grids: A Case Study on Short-Term Voltage Stability Assessment," *IEEE Trans. Ind. Inform.*, vol. 16, no. 3, pp. 1470–1482, Mar. 2020.

[6] N. Zhang, Y. Liu, F. Si, Q. Hou, A. Botterud, and C. Kang, "Topology and admittance estimation: Precision limits and algorithms," *iEnergy*, vol. 2, no. 4, pp. 297–307, Dec. 2023.

[7] Q. Zhao, W. Zhang, Z. Li, T. Zhang, J. Chen, and Y. Liu, "Robust H∞ State Estimation for Distribution System Considering Randomly Missing Measurements," *IEEE Trans. Instrum. Meas.*, vol. 72, pp. 1–16, 2023.

[8] S. K. Kotha, B. Rajpathak, M. Mallareddy, and R. Bhuvanagiri, "Wide area measurement systems based Power System State Estimation using a Robust Linear-Weighted Least Square method," *Energy Rep.*, vol. 9, pp. 23–32, Oct. 2023.

[9] Y. Zhang, Y. Xu, and Z. Y. Dong, "Robust Ensemble Data Analytics for Incomplete PMU Measurements-Based Power System Stability Assessment," *IEEE Trans. Power Syst.*, vol. 33, no. 1, pp. 1124–1126, Jan. 2018.

[10] Y. Zhang, Y. Xu, R. Zhang, and Z. Y. Dong, "A Missing-Data Tolerant Method for Data-Driven Short-Term Voltage Stability Assessment of Power Systems," *IEEE Trans. Smart Grid*, vol. 10, no. 5, pp. 5663–5674, Sept. 2019.

[11] M. He, V. Vittal, and J. Zhang, "Online dynamic security assessment with missing pmu measurements: A data mining approach," *IEEE Trans. Power Syst.*, vol. 28, no. 2, pp. 1969–1977, May 2013.

[12] T. Wu, Y.-J. A. Zhang, Y. Liu, W. C. Lau, and H. Xu, "Missing Data Recovery in Large Power Systems Using Network Embedding," *IEEE Trans. Smart Grid*, vol. 12, no. 1, pp. 680–691, Jan. 2021.

[13] A. Sagan, Y. Liu, and A. Bernstein, "Decentralized Low-Rank State Estimation for Power Distribution Systems," *IEEE Trans. Smart Grid*, vol. 12, no. 4, pp. 3097–3106, July 2021.

[14] M. Liao, D. Shi, Z. Yu, Z. Yi, Z. Wang, and Y. Xiang, "An Alternating Direction Method of Multipliers Based Approach for PMU Data Recovery," *IEEE Trans. Smart Grid*, vol. 10, no. 4, pp. 4554–4565, July 2019.

[15] S. Biswas and V. A. Centeno, "A Model-Agnostic Method for PMU Data Recovery Using Optimal Singular Value Thresholding," *IEEE Trans. Power Deliv.*, vol. 37, no. 4, pp. 3302–3312, Aug. 2022.

[16] J. Pei, Z. Wu, J. Wang, and D. Shi, "Robust Local PMU Measurement Recovery Based on Singular Spectrum Analysis of Hankel Structures," in *2021 IEEE Sustainable Power and Energy Conference (iSPEC)*, Dec. 2021, pp. 4233–4239.

[17] M. Yu, A. Neubauer, S. Brandt, and M. Kriegel, "TCN-BiLSTM-CE: An interdisciplinary approach for missing energy data imputation by contextual inference," *Appl. Energy*, vol. 401, p. 126618, Dec. 2025.

[18] J. J. Q. Yu, A. Y. S. Lam, D. J. Hill, Y. Hou, and V. O. K. Li, "Delay Aware Power System Synchrophasor Recovery and Prediction Framework," *IEEE Trans. Smart Grid*, vol. 10, no. 4, pp. 3732–3742, July 2019.

[19] L. Zhu and J. Lin, "Learning Spatiotemporal Correlations for Missing Noisy PMU Data Correction in Smart Grid," *IEEE Internet Things J.*, vol. 8, no. 9, pp. 7589–7599, May 2021.

[20] M. Zhang and Y. Chen, "Inductive Matrix Completion Based on Graph Neural Networks," in *Proc. ICLR*, vol. 3, no9. pp.1655-1669, Nov. 2015.

[21] J. Fang, L. Zheng, and C. Liu, "A Novel Method for Missing Data Reconstruction in Smart Grid Using Generative Adversarial Networks," *IEEE Trans. Ind. Inform.*, vol. 20, no. 3, pp. 4408–4417, Mar. 2024.

[22] Q.-H. Ngo, B. L. H. Nguyen, T. V. Vu, J. Zhang, and T. Ngo, "Physics-informed graphical neural network for power system state estimation," *Appl. Energy*, vol. 358, p. 122602, Mar. 2024.

[23] C. Ren and Y. Xu, "A Fully Data-Driven Method Based on Generative Adversarial Networks for Power System Dynamic Security Assessment With Missing Data," *IEEE Trans. Power Syst.*, vol. 34, no. 6, pp. 5044–5052, Nov. 2019.

[24] K. R. Mestav, J. Luengo-Rozas, and L. Tong, "Bayesian State Estimation for Unobservable Distribution Systems via Deep Learning," *IEEE Trans. Power Syst.*, vol. 34, no. 6, pp. 4910–4920, Nov. 2019.

[25] D. Xu, Z. Wu, J. Xu, Y. Zhu, and Q. Hu, "A Multiarea Forecasting-Aided State Estimation Strategy for Unbalance Distribution Networks," *IEEE Trans. Ind. Inform.*, vol. 20, no. 1, pp. 806–814, Jan. 2024.

[26] Y. Weng, R. Negi, C. Faloutsos, and M. D. Ilić, "Robust Data-Driven State Estimation for Smart Grid," *IEEE Trans. Smart Grid*, vol. 8, no. 4, pp. 1956–1967, July 2017.

[27] S. Moshtagh, B. Azimian, M. Golgol, and A. Pal, "Topology-Aware Graph Neural Network-Based State Estimation for PMU-Unobservable Power Systems," *IEEE Trans. Power Syst.*, vol. 40, no. 6, pp. 4547–4560, Nov. 2025.

[28] H. Li, Z. Ma, and Y. Weng, "A Transfer Learning Framework for Power System Event Identification," *IEEE Trans. Power Syst.*, vol. 37, no. 6, pp. 4424–4435, Nov. 2022.

[29] Y. Miyata and H. Ishikawa, "Handling concept drift in data-oriented power grid operations," *Meas. Energy*, vol. 7, p. 100052, Sept. 2025.

[30] J. Feng, Y. Chen, F. Li, A. Sarkar, and M. Zhang, "How Powerful are K-hop Message Passing Graph Neural Networks," in *Advances in Neural Information Processing Systems*, Inc., 2022, pp. 4776–4790.

[31] C. Genes, I. Esnaola, S. M. Perlaza, L. F. Ochoa, and D. Coca, "Robust Recovery of Missing Data in Electricity Distribution Systems," *IEEE Trans. Smart Grid*, vol. 10, no. 4, pp. 4057–4067, July 2019.

[32] J. Jiang et al., "ForkMerge: mitigating negative transfer in auxiliary-task learning," in *Proceedings of the 37th International Conference on Neural Information Processing Systems*, in NIPS '23. Red Hook, NY, USA: Curran Associates Inc., Dec. 2023, pp. 30367–30389.

[33] B. Yu, H. Yin, and Z. Zhu, "Spatio-temporal graph convolutional networks: a deep learning framework for traffic forecasting," in *Proceedings of the 27th International Joint Conference on Artificial Intelligence*, in IJCAI'18. Stockholm, Sweden: AAAI Press, July 2018, pp. 3634–3640.

[34] R. T. Dabou, I. Kamwa, A. Delavari, and F. S. Hasnaoui, "Big-Data Modeling Based Simscape Power Systems-ST for Protective Relaying," in *2025 15th International Conference on Power, Energy, and Electrical Engineering (CPEEE)*, Feb. 2025, pp. 257–262.

[35] N. H. A. Rahman and A. F. Zobaa, "Optimal PMU placement using topology transformation method in power systems," *J. Adv. Res.*, vol. 7, no. 5, pp. 625–634, Sept. 2016.

[36] J. H. Chow and K. W. Cheung, "A toolbox for power system dynamics and control engineering education and research," *IEEE Trans. Power Syst.*, vol. 7, no. 4, pp. 1559–1564, Nov. 1992.

[37] S. Konstantinopoulos, G. M. De Mijolla, J. H. Chow, H. Lev-Ari, and M. Wang, "Synchrophasor Missing Data Recovery via Data-Driven Filtering," *IEEE Trans. Smart Grid*, vol. 11, no. 5, pp. 4321–4330, Sept. 2020.

[38] X. Guo, S. Zhu, Z. Yang, H. Liu, and T. Bi, "Consecutive Missing Data Recovery Method Based on Long-Short Term Memory Network," in *2021 3rd Asia Energy and Electrical Engineering Symposium (AEEES)*, Mar. 2021, pp. 988–992.

[39] R. Bahmani and M. Afrasiabi, "Noisy PMU Data Recovery in Transient Conditions through Self-Attention Neural Networks," in *2024 IEEE PES Innovative Smart Grid Technologies Europe (ISGT EUROPE)*, Oct. 2024, pp. 1–5.

[40] X. Wang, R. Girshick, A. Gupta, and K. He, "Non-local Neural Networks," in *2018 IEEE/CVF Conference on Computer Vision and Pattern Recognition*, June 2018, pp. 7794–7803.